\documentstyle[aps, epsf, amsfonts]{revtex}
\begin{document}
\twocolumn[\hsize\textwidth\columnwidth\hsize\csname
@twocolumnfalse\endcsname
\date{\today}
\title{Cosmological perturbations in the 5D Big Bang.}
\author{Jaume Garriga$^{1,2,3}$ and Takahiro
Tanaka$^{3}$}
\address{$^1$ Departament de F{\'\i}sica Fonamental, Universitat
de Barcelona, Diagonal 647, 08028 Barcelona, Spain}
\address{$^2$ IFAE, Universitat Aut{\`o}noma
de Barcelona, 08193 Bellaterra (Barcelona), Spain}
\address{$^3$ Yukawa Institute for Theoretical Physics, Kyoto University,
Kyoto 606-8502, Japan}

\maketitle

\begin{abstract}

Bucher \cite{Bucher2001} has recently proposed an interesting 
brane-world cosmological
scenario where the ``Big Bang'' hypersurface is the 
locus of collision of two vacuum
bubbles which nucleate in a five dimensional flat space. 
This gives rise to an open universe, where the curvature 
can be very small provided that $d/R_0$ is sufficiently large.
Here, $d$ is the distance between bubbles and $R_0$ is their size
at the time of nucleation.
Quantum fluctuations develop on the bubbles
as they expand towards each other, and these in turn imprint
cosmological perturbations on the initial hypersurface. 
We present a simple formalism for calculating the
spectrum of such perturbations and their subsequent evolution. 
We conclude that, unfortunately, the spectrum is very tilted,
with spectral index $n_s=3$. The amplitude of fluctuations
at horizon crossing is given by $<(\delta \rho/\rho)^2> \sim
(R_0/d)^2 S_E^{-1} k^2$, where $S_E\gg 1$ is
the Euclidean action of the instanton describing the nucleation of a
bubble and $k$ is the wavenumber in units of the curvature scale. The 
spectrum peaks on the smallest possible relevant scale,
whose wave-number is given by $k\sim d/R_0$. We comment on the 
possible extension of our formalism to more general situations where 
a Big Bang is ignited through the collision of 4D extended objects.
\vskip .5 truecm

\end{abstract}

]

\vskip .5 truecm

\section{Introduction}

The possibility of having gravity localized \cite{RS,DGP} 
on a brane which moves in a higher dimensional space has
recently stimulated the search for alternatives to the standard inflationary 
paradigm \cite{ekpy,Bucher2001,git}. In these alternative scenarios, 
the Big Bang would result from the collision of 4-dimensional 
extended objects propagating in five dimensions. Pioneering
work in this direction \cite{ekpy} did not use the localization of gravity 
as an essential ingredient, but this was at the expense of a
somewhat singular behaviour at the moment of collision
\cite{new}, when the ``bulk'' in which the branes propagate
momentarily disappears.

The generic predictions of inflation seem to be in good agreement with 
current cosmological data, and therefore any alternative proposal 
has to measure up to high standards.
In particular, it is important to clarify the mechanisms by which
cosmological perturbations are generated, since at the moment these 
provide the finest tests for any theory of initial conditions.
The purpose of this paper is to consider this problem in the context of
the model proposed in Ref.\cite{Bucher2001}, where it seems to be best
posed (see also \cite{bpb}).

The paper is organized as follows. In Section II we briefly review Bucher's
scenario, describing the geometry of the problem. In
Section III we derive the expression for the primordial 
cosmological perturbations in terms of the physical parameters of the
model. In Section IV we discuss the subsequent evolution of such 
perturbations, and in Section V we summarize our conclusions.

\section{A Big Bang from bubble collision}

In Bucher's model one starts with a metastable (or ``false'') vacuum 
in five dimensions,
which is flat space or a very mildly expanding de Sitter space. This
decays through bubble 
nucleation into an anti-de Sitter (AdS) phase, where the five
dimensional cosmological constant is negative. 
The model requires the existence of degenerate discrete AdS vacua.
Upon collision of two bubbles corresponding to different vacua,
a domain wall forms, where gravity is localized
\`a la Randall-Sundrum \cite{RS}. 
The ``Big Bang'' hypersurface is the locus where the worldsheets of the 
two bubbles meet, and the domain wall, or ``local
brane'', is the place where we are supposed to live.

Let us use coordinates $X^A=(X^i,W,T)$ in the original 5D Minkowski
space, where a pair of bubbles of radius $R_0$ nucleate at $T=0$,
separated by a distance $2d$. The radius $R_0$ is a fixed parameter of
the theory, which is related to the bubble wall tension $\sigma$ and to the
energy gap $\epsilon$ between the original Minkowski phase and the
final AdS phases \cite{coleman} (this relation, however, will 
not be needed in the following discussion.)
Let us choose the origin of coordinates
to be at the center of one of the bubbles. After nucleation, the
bubble wall expands with constant acceleration, following a
hyperbolic trajectory which can be parametrized as
\begin{eqnarray}
T=R_0 \sinh\beta ,\\
\label{paramtraj}
R=R_0\cosh\beta,
\end{eqnarray}
where $R \equiv (|\vec X|^2+W^2)^{1/2}$, and
$\beta$ is the boost parameter of the bubble wall
$$
v=dR/dT=\tanh\beta.
$$
The collision takes place on the plane $W=d$, along the 
hyperboloid
\begin{equation}
T^2-|\vec X|^2 = d^2-R_0^2.
\label{hyper}
\end{equation}
To the future of this hyperboloid, and on the plane $W=d$, a
domain wall forms
where gravity is four dimensional \cite{RS,rs}.
Throughout this paper, we shall asume that, locally, the collision
process is almost instantaneous compared with the lengthscales of our 
interest, and
that a fixed fraction of the energy of the collision is channeled
into the degrees of freedom which live on the local brane.
We shall also assume the usual fine tuning between the tension of the
local brane and the AdS radius, so that the effective 4D cosmological 
constant is sufficiently small.

In the unperturbed setup, the 4D Minkowski space on the plane $W=d$ is 
matched into an open FRW
model along the hypersurface (\ref{hyper}). 
Along this hyperboloid, the metric is continuous, although its
derivative is not (this jump in the extrinsic curvature corresponds
to the jump in the Hubble rate accross the surface).
It is convenient to use Milne coordinates to describe the
4D Minkowski space before the collision, 
\begin{equation}
^{(4)}ds^2 = -dt^2 + t^2 d\Omega_H^2,
\label{milne}
\end{equation}
where the $t=const.$ surfaces have the hyperbolic geometry described
by $d\Omega_H^2$. In terms of Minkowski coordinates, Milne time is given by
$t=(T^2-|\vec X|^2)^{1/2}$. This form of the metric is valid for 
$t<R_0 \cosh^{-1}(d/R_0)$. 

For $t>R_0 \cosh^{-1}(d/R_0)$, the 4D metric
on the ``local brane'' is given 
by 
$$
^{(4)}ds^2 = -dt^2 + a^2(t) d\Omega_H^2.
$$
The local brane separates the
interiors of the two vacuum bubbles, which have already collided and
continue to expand. Due to the $O(3,1)$ symmetry of the process, 
the bulk metric
on both sides of the local brane is given by Schwarzschild-AdS 
\cite{rs,ruth}. The
evolution of the scale factor after collision is given by
the usual Friedmann equation
\begin{equation}
H^2 \approx {\kappa^2\over 3} \rho +{1\over a^2}.
\label{friedmann}
\end{equation}
Here, $H=\dot a/a$, $\kappa^2=8\pi G$ is the effective 
4D gravitational coupling and
$\rho$ is the matter energy
density which is deposited on the local brane after collision. This
will also include some ``dark radiation''\cite{rs}, because a
significant fraction of the
energy of the colliding bubbles may not stick to the brane; it may
just fly into the bulk
contributing to the effective mass of the Schwarzschild-AdS.
In (\ref{friedmann}) we have also neglected ``brany'' corrections 
in the right hand side, proportional to $\kappa^4 \rho^2 \ell^2$. 
These will be unimportant provided that the AdS radius $\ell$ in the 
bulk is sufficiently small. The discussion
of cosmological perturbations with these corrections included is 
significantly more complicated and it is left for future research.

From (\ref{hyper}), the spatial curvature scale of the
open universe at the moment of collision is given by
$(d^2-R_0^2)^{1/2}$, so the
flatness problem is solved provided that $d$ is sufficiently
large (as discussed in Section III.A, this gives a lower bound on
$d/R_0$). Of course, the homogeneity problem is also solved, due to the 
residual $O(3,1)$ symmetry of the colliding bubble setup along the 
hyperboloid (\ref{hyper}). Thus, in principle, Bucher's scenario 
seems to provide an attractive starting point for a brane-world 
cosmology.

\section{Initial perturbations}

Aside from solving the homogeneity and flatness problems, a serious 
candidate for a theory of initial conditions should explain the 
origin of the primordial cosmological perturbations. In brane
collision scenarios these may be seeded by preexisting fluctuations in
the shape of the colliding branes. These produce distortions of the 
``Big Bang'' hypersurface as well as initial perturbations in the distribution 
of the energy density. In this
Section we estimate this effect for the case of
colliding bubbles. 

Quantum fluctuations on expanding vacuum bubbles have been studied in 
Refs. \cite{gv}. For a bubble which is
centered at the origin of coordinates, the perturbed worldsheet 
$\tilde X^A$ is 
conveniently parametrized as  
$$
\tilde X^A(\xi^{\mu})= X^A(\xi^{\mu}) + N^A(\xi^{\mu}) \chi(\xi^{\mu})
=(1+\chi/R_0) X^A.
$$
Here, $N^A$ is the unit normal to the unperturbed worldsheet of the bubble,
which has the internal
geometry of a 4D de Sitter space of radius $R_0$, 
$\xi^{\mu}$ are a set of coordinates in this space and $X^A$ stands
for the unperturbed worldsheet (\ref{paramtraj}). It is known
\cite{gv} that the normal displacement $\chi$ behaves like a 
worldsheet scalar field with the tachyonic mass $m^2= -4 R_0^{-2}$,
which obeys an equation of the form
\footnote{Here, and in the following discussion, we are neglecting the 
self-gravity effect of the bubble walls on the perturbations. 
Domain walls produce a repulsive constant gravitational force
which causes an acceleration of order $(G\sigma)^{-1}$, where $G$ is the
five dimensional Newton's constant and $\sigma$ is the wall tension. 
If this length is very large compared with the inverse of the
proper acceleration of the bubble wall, given by $R_0^{-1}$, then the 
self gravity of the wall will be negligible.}
\begin{equation}
-\Box\chi - 4 R_0^{-2}\chi =0.
\label{diffeq}
\end{equation}
Here, $\Box$ is the covariant d'Alembertian in a 4D de Sitter space.

We must also consider the second bubble, whose center is at
a distance $2d$ from the first. For perturbations which are symmetric with
respect to the plane $W=d$, the collision will
still take place on this plane, the second bubble being just a
mirror image of the first. 
In what follows, we shall restrict our attention to such ``$Z_2$-
symmetric''perturbations. It is easy to show, following a calculation
similar to the one presented below, that the antisymmetric
mode has an amplitude comparable to that of the symmetric mode, but does not 
contribute to cosmological perturbations to linear order, so its
effect seems to be much smaller than that of the symmetric mode.

An important observation is that, by conservation of momentum, 
the fluid lines will be
orthogonal to the perturbed surface of collision, which is therefore
a co-moving, or velocity orthogonal surface\cite{LMW}. 
It is known (see next section)
that the {\em growing} mode of the curvature perturbation ${\cal R}_c$ on 
co-moving surfaces is a constant of motion on scales larger than the
Hubble rate, and hence ${\cal R}_c$ will be a quantity of interest to us.
In the present case, however, it will be equally important to consider
the $decaying$ mode, since its initial amplitude is comparable
to that of the growing mode. Therefore in order to characterize the
initial perturbation we will also need the density
perturbation $\delta_c$ on the same initial surface.

\subsection{Initial value of ${\cal R}_c$}

Due to local shifts $\delta t$ in the time of collision caused by 
fluctuations of the bubble shape, the Big Bang surface will
no longer be the smooth hyperboloid (\ref{hyper}). As
explained in the preceeding paragraphs, we can restrict attention to 
perturbations which are symmetric with respect to the plane $W=d$. 
Then, to linear order in $\delta t$, the metric induced on the Big Bang
surface is easily obtained from (\ref{milne}), and it is given by
$$
^{(3)}ds^2 = \left(1 +2{\delta t\over t}\right) t^2 d\Omega_H^2.
$$
Since we live in an almost flat universe, we shall be
interested in lengthscales which are short compared to the curvature
scale of the unperturbed surface. On these scales $d\Omega^H$ can be
replaced with a flat metric. Then, the intrinsic curvature $^{(3)}R$
of the perturbed surface is easily found to be
$$
^{(3)}R= {4 \over t^2}\Delta{\cal R}_c =
{4 \over t^2}\Delta {\delta t \over t}, 
$$
where $\Delta$ is the co-moving Laplacian.
The first equality is just the conventional definition of the
curvature perturbation used by most authors \cite{reviews}. In a flat
FRW with arbitrary scale factor and in an arbitrary gauge this definition
takes the form
\begin{equation}
^{(3)}R\equiv{4 \over a^2}\Delta{\cal R},
\label{rcdef}
\end{equation}
where now $^{(3)}R$ is
the perturbation of the intrinsic curvature scalar in the constant 
time surfaces of the corresponding gauge. 

To proceed, we must find $\delta t$ as a function of the normal
displacement $\chi$. Since all points on the unperturbed hyperboloid
are equivalent, we may just consider the vicinity of the point $\vec
X=0$, where $t=T$. It should be noted that the field $\chi$ lives on
the unperturbed bubble, which hits the collision plane at some value
$\beta_d$ of the unperturbed boost parameter determined by 
$
d= R_0 \cosh\beta_d.
$
However, the actual collision
does not take place at that time, but at
the time when the perturbed bubble reaches $W=d$, at some
value $\beta_c$ of the unperturbed boost parameter given by
$
d= (R_0 +\chi) \cosh\beta_c.
$ 
Thus, the shift in the time of collision is given by
\begin{eqnarray*}
\delta t =  \delta T &=& \tilde T - T \cr
&=& (R_0+\chi) \sinh\beta_c - R_0 \sinh\beta_d
\approx -{\chi \over \sinh\beta_d}.
\end{eqnarray*}
We conclude that the curvature perturbation on the initial co-moving
surface is given by
\begin{equation}
{\cal R}_{c(i)} = {\delta T\over T} = -{\chi \over R_0 \sinh^2
\beta_d}.
\label{Rc}
\end{equation}
Let us now estimate the size of this effect.

As mentioned above, $\chi$ satisfies the equation for a scalar field
on the worldsheet de Sitter space, Eq. (\ref{diffeq}). The
corresponding  canonical field $\phi$ 
with dimensions of mass is related to $\chi$ by \cite{gv}
$$
\phi = \sigma^{1/2} \chi,
$$
where $\sigma$ is the tension of the wall.
As is well known, when a given mode of a nearly massless scalar field 
crosses the de Sitter horizon 
$R_0^{-1}$, it ``freezes'' with some amplitude $\phi_k \sim
R_0^{-1}$.
However, in our case this amplitude does not stay constant. 
Due to the tachyonic mass $m^2 = -4 H^2$ in (\ref{diffeq}), a mode
with wavenumber $k$ grows proportionally to $e^{\beta}$ as
\begin{equation}
\chi_k(\beta)\sim \sigma^{-1/2}R_0^{-1} e^{\beta-\beta_k},
\label{ref}
\end{equation}
where $\beta_k$ is the value of the time-like boost parameter [see
Eq. (\ref{paramtraj})] at which
the physical wavelength of the mode $k^{-1}R_0 e^{\beta}$ becomes larger
than the inverse expansion rate on the bubble, $R_0$. 
Here we adopt
the convention that the co-moving curvature scale corresponds to 
$k\sim 1$. 
It is easy to 
see that at the time of collision we have 
$\exp(\beta_d-\beta_k)\sim k^{-1} (d/R_0)$, 
and therefore, 
$$
\chi_k(\beta_c)\sim \sigma^{-1/2}{d\over R_0^2}k^{-1}.
$$
Using $\exp{\beta_d}\approx d/R_0$ in Eq. (\ref{Rc}) we find that
by order of magnitude, 
\begin{equation}
{\cal R}_{c(i)}(k) \sim \left({1\over S_E}\right)^{1/2}{R_0\over d}\, k^{-1}.
\label{Rcestimate}
\end{equation}
Here $S_E \sim \sigma
R_0^4 \gg 1$ is the Euclidean action of the instanton describing the
nucleation of the bubble.

This perturbation is rather minute even on scales comparable to the 
curvature scale $k\sim 1$. An upper bound on $R_0/d$ can be obtained 
as follows.
The spatial curvature radius of our present universe is much larger
than the present Hubble radius $H_0^{-1}$. 
Taking into account that the curvature
radius of the initial surface was given by $d$, this
leads to the constraint
$$
z_i d \gtrsim H_0^{-1},
$$
where $z_i$ is the redshift at which the collision surface is. This
redshift is given by
$$
z_i \approx \left({\rho_i\over \rho_0}\right)^{1/4} \lesssim 
\left(\sigma {d\over R_0} {\kappa^2 \over H_0^2}\right)^{1/4},
$$
where we assume radiation dominance for most of the cosmological evolution.
In the last inequality we have used the fact that the energy density after
collision will not be larger than the input energy density, which is
of order $\sigma \cosh\beta_d$. From the
two previous inequalities we find
$$
{R_0\over d} \lesssim 10^{-24} S_E^{1/5}.
$$
Therefore, we
have
\begin{equation}
{\cal R}_{c(i)}(k) \lesssim 10^{-24} S_E^{-3/10} k^{-1}.
\label{bound}
\end{equation}
If there was no initial density perturbation, then this quantity would
stay nearly constant until horizon reentry (see next section). 
This would lead to an amplitude of perturbations smaller than the
observed value by some 20 orders of magnitude. As we shall see, the
bound (\ref{bound}) still applies to scales comparable to the
curvature scale. However, on smaller scales the initial density perturbation
$\delta_{c(i)}$ induces a much larger perturbation on
${\cal R}_c$ by the time of horizon reentry. Unfortunately, this
comes at the prize of a very tilted spectrum, which is not compatible
with observations.

\subsection{Initial value of $\delta_c$}

To evolve the cosmological perturbations, two initial
conditions are needed for each wavelength. In addition 
to the initial curvature 
${\cal R}_c$ we will also need the density perturbation $\delta_c$. 
By assumption, a fixed fraction of the energy of the
collision goes into the brane, and therefore the density at the moment of
collision is proportional to the Lorentz factor $\gamma
=\cosh\tilde\beta$. 
Since the 
boost parameter is additive, the perturbed one will be given by 
$$
\tilde\beta=\beta_c+\dot \chi,
$$
where
$$
\dot\chi = {1\over R_0} {\partial\chi \over \partial\beta} 
$$
is the derivative of the perturbation with respect to the proper time
$\tau=R_0\beta$ measured by an observer on the bubble. 
Thus, the change in the
boost parameter is given to linear order by
$$
\delta\beta=\tilde\beta-\beta_d = \dot\chi-{\chi \over
R_0}\coth\beta_d.
$$
With these relations, we
obtain the density perturbation at the moment of collision as
\begin{equation}
\delta_{c(i)} = \tanh\beta_d \delta\beta = \dot\chi \tanh\beta_d -
{\chi\over R_0}.
\label{deltac}
\end{equation}
As noted above, in the case of our interest, the r.m.s. fluctuation in 
$\chi$ grows exponentially
fast with $\beta$. However, the combination which enters $\delta_c$
anihilates the leading term in this exponential dependence, and leaves
only a contribution which decays with $\beta$. From this argument
alone, it should be clear that the spectrum of $\delta_c$ will not be scale
invariant, and that it will have opposite tilt to the spectrum of
${\cal R}_c$. To extract the spectral index, we need to look at the 
detailed form of the mode functions.

For definiteness, we shall use the open chart on the de Sitter
worldsheet of the bubble. The conclusion, however, is independent of
our slicing since we are interested only in wavelenths much smaller
than the curvature scale. In this chart, equation (\ref{diffeq})
for the evolution of
$\chi$ reads
\begin{equation}
\ddot \chi_k +{3\over R_0}\coth\beta\dot\chi_k -
{4\over R_0^2}\chi_k + {1+k^2 \over R_0^2
\sinh^2\beta}\chi_k=0.
\label{modequat}
\end{equation}
The quantum state of perturbations on a nucleating bubble is uniquely
determined by de Sitter invariance \cite{vv}. It is given by the so called
Bunch-Davies vacuum (also known as Euclidean vacuum). The
corresponding modes in the open chart have been studied by
\cite{st,bgt}, 
and for a scalar field with mass $m^2=-4R_0^2$ they 
are given by
$$
\chi_k(\beta)\propto {P_2^{ik}(\cosh\beta)\over \sinh\beta},
$$
where $P_2^{ik}$ are the Legendre functions with the branch cut from
$-\infty$ to $1$ on the real axis. In the above equation we have 
ignored a contribution proportional to the decaying mode, which is 
accompanied by the factor $e^{-\pi k}$ and which is therefore
irrelevant at wavelenths much shorter than the curvature scale.
Expanding the Legendre function for large $\beta$, we have
\begin{eqnarray}
\chi_k(\beta)\sim {\sigma^{-1/2}R_0^{-1}\over \sinh\beta_k}
\left[\sinh\beta +{k^2+4\over 6 \sinh\beta} +O(e^{-3\beta})\right].
\label{ref2}
\end{eqnarray}
Here, the normalization is fixed as in Eq.(\ref{ref}), the only
difference being that we have been careful to keep more terms in the
expansion because the leading one clearly does not contribute to 
(\ref{deltac}). Substituting (\ref{ref2}) in (\ref{deltac}) and using
$\exp\beta\approx d/R_0$ we have
\begin{equation}
\delta_{c(i)} \sim\left({1\over S_E}\right)^{1/2}{R_0\over d}\, k.
\label{deltacestimate}
\end{equation}
For $k\sim 1$, this is of the same order as the curvature
perturbation given in (\ref{Rcestimate}), where both are very small.
However, (\ref{deltacestimate}) can be much larger on small scales.
Unfortunately, this is due to a strong tilt in the power spectrum,
corresponding to $n_s=3$ in the standard notation.

\section{Evolution of the perturbations}

Eqs. (\ref{Rc}) and (\ref{deltac}) provide the initial conditions
(position and momentum, as it were) for the evolution of cosmological 
perturbations in the FRW phase. 
As mentioned above, since our universe is reasonably flat, the scales
of interest to us will be much smaller than the curvature scale and it
is a good approximation to consider perturbations to a flat FRW
universe. 

In an arbitrary gauge, the perturbed metric for scalar perturbations
can be written as \cite{reviews}
\begin{eqnarray*}
 d\tilde s^2=a^2(\eta)^2\Biggl[
    &-&(1+2A Y) d\eta^2-2 B Y_i d\eta dx^i 
\cr
     && +\left[(1+2D Y)\delta_{ij}+2E Y_{ij}\right]dx^i dx^j\Biggr],  
\end{eqnarray*}
where, $Y \propto e^{i{\bf kx}} $ is the appropriately 
normalized plane wave, 
and summation over modes with different wave number ${\bf k}$ 
is omitted.  
The vector and the traceless tensor constructed from $Y$ are
defined by 
$
 Y_i = -k^{-1} Y_{,j},$
and
$
Y_{ij}= k^{-2} Y_{,ij}+{1\over 3}\delta_{ij} Y. 
$
We also use convenient combinations of metric perturbations 
defined by 
\begin{eqnarray}
 && {\cal R}=D+{1\over 3}E,
\cr
 && k \sigma_g =E'-k B,
\label{defR}
\end{eqnarray}
where a prime denotes differentiation with respect to $\eta$.  
${\cal R}$ is the quantity related to the perturbation of the 
spatial scalar curvature of a time slice through Eq. (\ref{rcdef}),
and $\sigma_g$ is related to the shear of the hypersurface normal
vector field~\footnote{For the present discussion, the geometric
interpretation of $\sigma_g$ will be irrelevant, it will just be used
as a convenient variable.}$\!\!$.
The familiar gravitational potential in the Newton gauge is 
gauge invariantly defined as 
\begin{equation}
 \Phi =-{\cal R}+{\cal H}k^{-1}\sigma_g,
\end{equation}
where ${\cal H}=a'/a$. 

The energy momentum tensor of a perturbed perfect fluid is given by 
\begin{equation}
 \tilde T^{\mu\nu}=(\tilde \rho+\tilde P)\tilde u^{\mu}
    \tilde u^{\nu}+\tilde P\tilde g^{\mu\nu}, 
\end{equation}
with
$\tilde\rho  =  \rho+ Y \delta\rho$,
$\tilde P =  P + Y \delta P$, 
$\tilde u^{0} =  a^{-1}(1-A Y),$ and 
$\tilde u^{i} = a^{-1} v Y^i.$
We shall assume that the fluid consists of a single component. 
Then the ratio between the density and pressure perturbations 
becomes a function of the background energy density. 
We denote this ratio by 
$c_s^2={\delta P/ \delta\rho}$. 

In the comoving gauge, in which $v-B=0$, the 
perturbed Einstein equations become
\begin{equation}
{\cal H} A ={\cal R}_c',
\label{15}
\end{equation}
and
\begin{eqnarray}
&& {\cal H} (k\sigma_g)-k^2{\cal R}_c=-{\kappa^2 a^2\over 2}\delta \rho,
\label{16a}
\\
&& (k\sigma_g)'+2{\cal H} (k\sigma_g) - k^2
    \left({{\cal R}_c'\over {\cal H}}+{\cal R}_c\right) 
      =0,
\label{16b}
\\
&& {{\cal H}'-{\cal H}^2\over {\cal H}}{\cal R}_c'
   = {\kappa^2 a^2\over 2}\delta P. 
\label{16c}
\end{eqnarray}
Eqs.(\ref{16a}), (\ref{16b}) and (\ref{16c}) have already been
simplified by using (\ref{15}). 
For reference, we also quote the background equations 
\begin{equation}
 {\cal H}'=-{\kappa^2 a^2\over 6}(\rho+3P),
\quad
 {\cal H}^2={\kappa^2 a^2\over 3} \rho. 
\end{equation}
Eqs. (\ref{15}) and (\ref{16a}) give the following expression for the gauge 
invariant $\Phi$ in terms of the density perturbation $\delta_c$
\begin{equation}
  \Phi=-{3{\cal H}^2\over 2k^2}{\delta_c}. 
\label{rel}
\end{equation}
The value of $\Phi$ is larger than $\delta_c$ by the 
factor of ${\cal H}^2/k^2$, and hence this is not a particularly
illuminating variable on very large scales.

The evolution equations for given initial values for 
${\cal R}_c$ and $\delta_c$ can be obtained by 
combining Eqs(\ref{16a}), (\ref{16b}) and (\ref{16c}): 
\begin{eqnarray} 
&& {d{\cal R}_c\over dN}= -c_s^2 {\delta_c \over (1+w)},
\label{17}\\
&&{{\cal H}\over a^2}{d\over dN}(a^2{\cal H}\delta_c)= k^2(1+w) 
{\cal R}_c,  
\label{18}
\end{eqnarray}
where $w=P/\rho$ is the parameter characterizing the equation of state
and  $dN={\cal H}d\eta$. 

To discuss the evolution of perturbations it is useful to 
formally integrate the second equation,
\begin{eqnarray}
\delta_c &=& {1\over a^2{\cal H}}\int_{N_i}^{N} (a^2{\cal H}) {k^2\over {\cal
H}^2}(1+w){\cal R}_c dN 
\cr
 &&+{1\over a^2{\cal H}}(a^2{\cal H}\delta_c)_{(i)}.
\label{integral}
\end{eqnarray}
The function $a^2/{\cal H}$ is an increasing function, 
so provided that ${\cal R}_c$ stays approximately constant at late times
(which will be a self-consistent assumption), 
the integral in the right hand side is dominated by 
the contribution from the neighborhood of the upper boundary 
of integration. 
If we consider a simple case in which 
$c_s^2=w=$const., the scale factor and ${\cal H}$ are 
given by 
\begin{equation}
 a=(\eta/\eta_0)^{2\over 1+3w}, 
\quad 
 {\cal H}={2\over 1+3w}\eta^{-1}. 
\end{equation}
Performing the integral in (\ref{integral}) we have
\begin{equation}
\delta_c \approx {2+2 w\over 5+3 w}{k^2\over {\cal H}^2}{\cal R}_c + 
{1\over a^2{\cal H}}(a^2{\cal H}\delta_c)_{(i)},
\label{int}
\end{equation}
where we have assumed ${\cal R}_c\approx const.$ at late times 
\footnote{Incidentally, even
when ${\cal R}_c$ is not constant, but increasing with time, the order of
magnitude of the first term in (\ref{int}) does not change.}$\!\!$.
Initially, ${\cal R}_c\lesssim \delta_c$ and consequently 
the first term in the right hand side is suppressed with respect
to the second at least by a factor of $(k^2/{\cal H}^2)\ll 1$.  
Substituting the dominant part into Eq.(\ref{17}), we have 
\begin{equation}
 {\cal R}_{c(f)}
   \approx {\cal R}_{c(i)}-
       (a^2{\cal H}\delta)_{(i)} \int_{\eta_i}^{\infty} 
                   {d\eta\over a^2}{c_s^2\over 1+w}. 
\label{R_cf}
\end{equation}
Then, this integration is performed to obtain  
\begin{equation}
 {\cal R}_{c(f)}
   \approx {\cal R}_{c(i)}
       -{2w \over 3(1-w^2)}\delta_{c(i)}. 
\label{final}
\end{equation}
In obtaining (\ref{final}) we have neglected the contribution from
the first term in the right hand side of (\ref{int}). It is easy to
check that this term does not have any effect until the wavelength of
the mode is comparable to the hubble radius $k \sim {\cal H}$.
At that time, the second term in (\ref{int}) is unimportant, and the
amplitude of density perturbations at horizon crossing can be read off
from the first term
\begin{equation}
\delta_c|_{hc}\approx {2+2 w\over 5+3 w}{k^2\over {\cal H}^2}{\cal R}_{c(f)}. 
\end{equation}
Finally, the Newtonian potential at horizon crossing can be found from
the relation (\ref{rel})
\begin{equation}
\Phi|_{hc} \approx -{3+3 w\over 5+3 w}{\cal R}_{c(f)}.
\end{equation}
Substituting (\ref{final}) into the previous two equations and using
(\ref{Rcestimate}) and (\ref{deltacestimate}) we have
$$
\Phi|_{hc}\sim \delta_c|_{hc} \sim {\cal R}_{c(f)} \sim 
{R_0\over d} S_E^{-1/2} k, 
$$
which of course displays the same strong spectral tilt as the initial
density perturbation $\delta_{c(i)}$.

\section{conclusions}

In this paper, we have discussed 
the primordial spectrum of density perturbations 
in the brane-world model proposed by Bucher\cite{Bucher2001}. In this 
scenario, two bubbles nucleate in the 5-dimensional Minkowski bulk, and 
their collision forms a brane where gravity is localized. 
This model solves the homogeneity and flatness problems, provided that 
the separation $2d$
of the bubble nucleation points is sufficiently large 
compared with the bubble radius $R_0$. 

We have evaluated the initial spectrum of scalar density perturbations,
assuming that they originate from quantum fluctuations on the bubbles.  
An important observation is that the surface of collision coincides with
the so-called comoving hypersurface (or velocity orthogonal slicing). 
On this initial surface, we first evaluated the density 
contrast $\delta_c=\delta\rho/\rho$ and the spatial curvature perturbation 
${\cal R}_c$ [See Eq.~(\ref{defR}) for definition]. 
The initial value $\delta_{c(i)}$ turns out to have a 
very steep spectrum corresponding to $n_s=3$, while 
the spectral index for ${\cal R}_{c(i)}$ is $n_s=-1$. The 
amplitude of both perturbations is comparable at wavelengths
of the order of the curvature scale, 
and therefore $\delta_{c(i)}$ has a larger amplitude on scales
relevant to present observations. 

It is known that the curvature perturbation in the 
comoving gauge, ${\cal R}_c$, is conserved on scales 
much larger than the horizon, and therefore
this quantity is often used to discuss the evolution of 
perturbations in the early universe. 
The constancy of ${\cal R}_c$, however, does not 
hold for the decaying mode. 
In the present case, the initial conditions which arise as 
a result of bubble collision, contain a significant 
amount of ``contamination'' from the decaying mode. 
Therefore, ${\cal R}_c$ does not stay constant   
in the subsequent evolution of perturbations, which 
is described by two coupled first order differential 
equations. 
Given initial values $\delta_{c(i)}$ and ${\cal R}_{c(i)}$,  
we solved these equations to find that the final 
value of  ${\cal R}_{c}$ (at the time of
horizon crossing) becomes comparable to the initial value of 
the density perturbation $\delta_{c(i)}$ 
[See Eq.~(\ref{R_cf})]. 
This final value of ${\cal R}_c$ also gives the order of magnitude
of the density contrast at horizon crossing. 
Thus, we conclude that the spectral index for 
primordial density perturbations is $n_s=3$. 
The estimated amplitude is very small at wavelengths comparable 
to the curvature scale. Perturbations of $O(10^{-5})$ at the 
present horizon scale may of course be obtained by choosing the 
ratio between the curvature scale and the present horizon scale 
appropriately, but the spectrum is too steep to be consistent with 
observations. 

Although in this paper we have investigated a particular realization
of the brane Big Bang, we may have learned a few lessons which may 
be useful in more general cases. 
First, in the present model the bubble fluctuations are 
described by an effective 4-dimensional scalar 
field with a negative mass squared. 
Consequently, fluctuations on the bubble worldsheet have a red
spectrum,  which has a larger amplitude for longer wavelengths. 
However, the resulting spectrum of density perturbations 
turned out to be a blue one. 
This means that in principle it may be possible (although perhaps not easy)
to generate a nearly scale invariant spectrum 
even when the mass of the effective field 
corresponding to the bubble fluctuations is not close to zero.
Second, we need to be careful 
in using ``standard'' results of cosmological perturbation theory, 
which in some cases neglect 
the contribution from the decaying mode. In order to determine the
evolution just after collision, two initial conditions must be
supplied, and both turn out to be important.

\section*{acknowledgements}

We are grateful to Jose Juan Blanco-Pillado and 
Martin Bucher for useful discussions.
J.G. is partially supported by the Templeton Foundation under grant 
COS 253, by CICYT under grant AEN98-0431 and by the Yamada
Foundation. T.T. is partially supported by the Monbukagakusho 
Grant-in-Aid No.~1270154, and by the Yamada Foundation.

\end{document}